\documentclass[11pt,twoside]{article}
\usepackage{./asp2014}

\aspSuppressVolSlug
\resetcounters

\begin{document}

\title{Blade Runner --What kind objects are there in the JVO ALMA Archive?--}
\author{Satoshi Eguchi,$^{1}$ Yuji Shirasaki,$^{2}$ Christopher Zapart,$^{2}$
	Masatoshi Ohishi,$^{2}$ Yoshihiko Mizumoto,$^{2}$ Wataru Kawasaki,$^{2}$
	Tsuyoshi Kobayashi,$^{2}$ George Kosugi$^{2}$}
\affil{$^{1}$Fukuoka University, Department of Applied Physics, Faculty of Science, Fukuoka, Japan; \email{satoshieguchi@fukuoka-u.ac.jp}}
\affil{$^{2}$National Astronomical Observatory of Japan, Osawa, Mitaka, Tokyo, Japan}

\paperauthor{Satoshi Eguchi}{satoshieguchi@fukuoka-u.ac.jp}{}{Fukuoka University}{Department of Applied Physics, Faculty of Science}{Fukuoka}{Fukuoka}{814-0180}{Japan}
\paperauthor{Yuji Shirasaki}{yuji.shirasaki@nao.ac.jp}{}{National Astronomical Observatory of Japan}{Astronomy Data Center}{Mitaka}{Tokyo}{181-8588}{Japan}
\paperauthor{Christopher Zapart}{chris.zapart@nao.ac.jp}{}{National Astronomical Observatory of Japan}{Astronomy Data Center}{Mitaka}{Tokyo}{181-8588}{Japan}
\paperauthor{Masatoshi Ohisi}{masatoshi.ohishi@nao.ac.jp}{}{National Astronomical Observatory of Japan}{Astronomy Data Center}{Mitaka}{Tokyo}{181-8588}{Japan}
\paperauthor{Yoshihiko Mizumoto}{mizumoto.y@nao.ac.jp}{}{National Astronomical Observatory of Japan}{Division of Optical and Infrared Astronomy}{Mitaka}{Tokyo}{181-8588}{Japan}
\paperauthor{Wataru Kawasaki}{wataru.kawasaki@nao.ac.jp}{}{National Astronomical Observatory of Japan}{NAOJ Chile Observatory (MITAKA)}{Mitaka}{Tokyo}{181-8588}{Japan}
\paperauthor{Tsuyoshi Kobayashi}{tsuyoshi.kobayashi@nao.ac.jp}{}{National Astronomical Observatory of Japan}{NAOJ Chile Observatory (MITAKA)}{Mitaka}{Tokyo}{181-8588}{Japan}
\paperauthor{George Kosugi}{george.kosugi@nao.ac.jp}{}{National Astronomical Observatory of Japan}{NAOJ Chile Observatory (MITAKA)}{Mitaka}{Tokyo}{181-8588}{Japan}

\begin{abstract}
The JVO ALMA Archive provides users one of the easiest ways to access the ALMA archival data.
The users can have a quick look at a 3 or 4-dimensional data cube without downloading
multiple huge tarballs from a science portal of ALMA Regional Centers (ARCs).
Since we just synchronize all datasets with those of ARCs,
the metadata are identical to the upstream, including ``target name'' for each dataset.
The name is not necessarily a common one like NGC numbers, but sometimes one of sequential
numbers assigned in an observation proposal.
Compilation of these artificial names into astronomical ones could provide users more flexible
and powerful search interfaces;
for instance, with the knowledge of the redshift for each source, the users can easily find
the datasets which observed their interested emission/absorption lines at not the observer
frame but the rest frame, fitting well with theoretical studies.
To implement this functionality, cross-identification of all the sources in our archive with
those in some other astronomical databases such as NED and SIMBAD is required.
We developed a tiny Java application named ``Blade Runner'' for this purpose.
The program works as a crawler for both the JVO ALMA Archive and SIMBAD, storing all information
onto a SQLite-based database file;
this portable design enables us to communicate results to each other even under different
computing environments.
In this paper, we introduce its software design and our recent work on the application,
and report a preliminary result on the source identification in our archive.
\end{abstract}

\section{Introduction}

The Atacama Large Millimeter/submillimeter Array (ALMA) is one of the largest radio
telescopes, built in Chile.
Thanks to its high spatial and frequency resolution, it continuously brings us new knowledge
of the universe.
All data obtained with ALMA are released to the public via the science portal of
ALMA Regional Centers (ARCs) known as the ALMA Science Archive after one-year
proprietary period from observations.
In the archive, calibrated data (measurement sets) and their FITS images are packed into
multiple GB-size \texttt{tar} files;
to quick look at these FITS images requires the users to download some (all in some cases) of
the huge tarballs.
For easy access to the FITS images, we provide a web interface named the JVO ALMA
Archive on the portal site of the Japanese Virtual Observatory (JVO) project \citep{Eguchi2014}.

In the JVO ALMA Archive, we fetch all tarballs from the upstream every day.
FITS images are extracted from the tarballs, and unique IDs called dataset IDs\footnote{One observation corresponds to multiple dataset IDs.}
are assigned to the images.
Metadata for each dataset are drawn from the FITS header and the readme file,
which holds a project code title\footnote{A project is an unique ID assigned to each observational proposal by ARCs,
	and a project code title is a summary of it.}, then the datasets are released to the public.
Each metadata contain the target position and ``name'', which is sometimes a common one
like NGC numbers but sometimes one of sequential numbers assigned in a proposal.
If we can compile such artificial names into astronomical ones and link them
to entries in astronomical databases such as SIMBAD and NED,
the users can utilize our archive more effectively.
In this paper, we report our recent work to recover true target names from the metadata.

\section{Strategies}

Our goal is to find counterparts for each ALMA source in astronomical databases.
We utilize SIMBAD for this task since it covers both galactic and extragalactic
sources.
For simplicity, we do not consider whether SIMBAD sources were truly detected
with ALMA, and just check position relationship between each others;
information about the target position, ``target name'', width and hight of
the FITS image, and project code title of the dataset is adequate for this purpose.

We create a source list whose sources locate within a 20$^{\prime \prime}$
(60$^{\prime \prime}$ in some diffuse ones) radius circle from the center
of each dataset by utilizing the coordinate search functionality of
SIMBAD, seeking for a nearest well-known source in the list.
If we find a candidate, then we check the consistency with the description
of the project code title and adopt the candidate as a counterpart if this is the case;
otherwise we apply the same criterion to the 1st, 2nd, $\cdots$ nearest object in sequence
until we find a proper counterpart or there is no source left in the list.
In the latter case, we identify that there is no counterpart for the dataset.

\section{Implementation}

Since we have to take project code titles, which are written in natural
language (English), into account, identification of a counterpart for each
dataset should be performed by hand.
In addition, we have to switch two databases, the JVO ALMA Archive and SIMBAD.
Hence we developed a tiny Java application named Blade Runner, which is
designed to assist a user to perform source identification by summarizing
all information from various databases in a screen.

This work is a collaboration between Fukuoka University and the JVO project
at National Astronomical Observatory of Japan (NAOJ), and the bandwidth of
the network between us is rather narrow.\footnote{The effective bandwidth between the two
institutes is about 20\,MB/s.}
To work together and share the results under limited network bandwidth,
we designed for Blade Runner to communicate only metadata of datasets, adopting
SQLite3 for its database engine since a database is stored in a single file and
we never access it simultaneously.

Blade Runner fetches the dataset information of the JVO ALMA Archive
by accessing the JVO portal site over the standard HTTP protocol,
and parsing the HTML\footnote{This information can also be obtained over the TAP protocol.}
with the jsoup library.\footnote{\url{http://jsoup.org/}}
Then the application crawls SIMBAD by querying sources in a 60$^{\prime \prime}$
radius circle from the center of each dataset with the coordinate search functionality;
the result returned by SIMBAD is formatted in a VOTable.

After synchronization of databases, Blade Runner shows all dataset IDs in a list.
Fig.~\ref{P035-figure-bladerunner} is a screen shot of Blade Runner.
Selecting one ID from the list displays the ``target name'', target position, project
code title, and a list of the counterpart candidates in a single window;
there is also a button to jump directly to the information page of the dataset
in the JVO ALMA Archive.
This simple GUI design enables the users to identify counterparts effectively.

\articlefigure{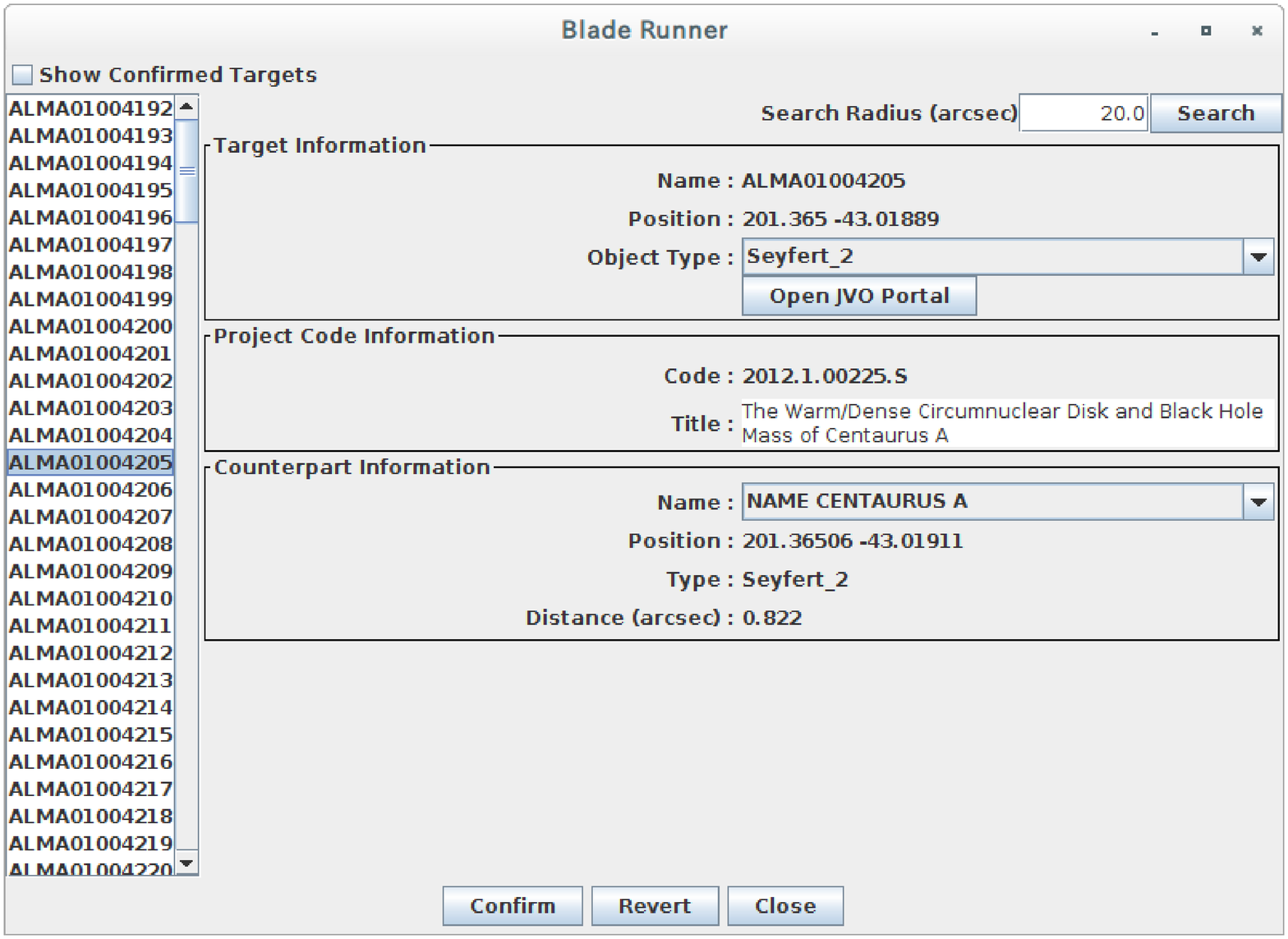}{P035-figure-bladerunner}{A screen shot of Blade Runner.}

\section{Results}

We applied Blade Runner to the ALMA JVO Archive on 2015 August 31.
There were 4,410 datasets in the archive.
We found that 934 SIMBAD sources have been observed uniquely with ALMA.
The categories of the sources based on the object type information in SIMBAD are
summarized in Table~\ref{P035-table-classification}, comparable to \citet{Meankins2014}.
The number of papers concerning extragalactic sources is rather smaller than
galactic ones in the literature, but more than 60\% of the JVO ALMA Archive sources
was found to be extragalactic; this suggests that our archive can be ``a gold mine''.
This classification result is available from the JVO portal site.

\begin{table}[!ht]
 \caption{Source classifications of the JVO ALMA Archive.\label{P035-table-classification}}
 \smallskip
 \begin{center}
  {\small
   \begin{tabular}{ccc}
    \tableline
    \noalign{\smallskip}
    Category & Number & Ratio (\%) \\
    \noalign{\smallskip}
    \tableline
    \noalign{\smallskip}
    Clusters of galaxies & 22 & 2 \\
    Galaxies & 395 & 42 \\
    LINERs, AGNs, blazars, QSOs, starburst galaxies & 180 & 19 \\
    Clusters of stars & 4 & $<$1 \\
    Stars, white dwarfs, neutron stars, black holes, brown dwarfs & 173 & 19 \\
    Young stellar objects & 30 & 3 \\
    Planets & 1 & $<$1 \\
    Molecular clouds, nebulae, supernova remnants, interstellar mediums & 54 & 6 \\
    Individual sources & 63 & 7 \\
    Unknowns & 1 & $<$1 \\
    \noalign{\smallskip}
    \tableline\ 
   \end{tabular}
  }
 \end{center}
\end{table}

\section{Future Work}

At this time, redshift information in SIMBAD is not crawled.
The knowledge of redshifts enables users to construct a query for searching
their interested emission/absorption lines from the rest frame wavelengths.

Blade Runner cannot handle multiple counterparts currently.
The users of our archive can like to search their interested astronomical sources
just based on the position relationships regardless of real detection with ALMA.
This extension can be implemented by utilizing crawled data of SIMBAD.

Blade Runner does not have a functionality to display a FITS image of each
dataset because of limited bandwidth between our institutes.
The JVO ALMA Archive has binned (compressed) FITS images internally \citep{Eguchi2014}.
Utilization of these images would enable the application to superpose positions
of SIMBAD sources onto the ALMA images like Aladin Lite even under such network environment.

\bibliography{P035}  

\end{document}